\documentclass[useAMS,usenatbib]{mn2e}
\usepackage{amsmath,graphicx,fleqn,txfonts}

\title{Phase-space mixing and the merging of cusps}

\author[Walter~Dehnen]
{
  Walter Dehnen\\
  Department of Physics \& Astronomy,
  University of Leicester,
  Leicester, LE1~7RH
}

\begin{document}

\maketitle

\begin{abstract}
  Collisionless stellar systems are driven towards equilibrium by mixing of
  phase-space elements. I show that the \emph{excess-mass function}
  $D(f)=\int_{\Bar{F}(\bmath{x},\bmath{v})>f} (\Bar{F}(\bmath{x}, \bmath{v})
  -f)\,{\mathrm{d}}^3\!\bmath{x}{\mathrm{d}}^3\!\bmath{v}$ (with $\Bar{F}
  (\bmath{x}, \bmath{v})$ the coarse-grained distribution function) always
  decreases on mixing. $D(f)$ gives the excess mass from values of
  $\Bar{F}(\bmath{x},\bmath{v})>f$.  This novel form of the mixing theorem
  extends the maximum phase-space density argument to all values of $f$. The
  excess-mass function can be computed from $N$-body simulations and is
  additive: the excess mass of a combination of non-overlapping systems is the
  sum of their individual $D(f)$. I propose a novel interpretation for the
  coarse-grained distribution function, which avoids conceptual problems with
  the mixing theorem.
  
  As an example application, I show that for self-gravitating cusps ($\rho
  \propto r^{-\gamma}$ as $r\to0$) the excess mass $D\propto
  f^{-2(3-\gamma)/(6-\gamma)}$ as $f\to\infty$, i.e.\ steeper cusps are less
  mixed than shallower ones, independent of the shape of surfaces of constant
  density or details of the distribution function (e.g.\ anisotropy).  This
  property, together with the additivity of $D(f)$ and the mixing theorem,
  implies that a merger remnant cannot have a cusp steeper than the steepest of
  its progenitors. Furthermore, I argue that the remnant's cusp should not be
  shallower either, implying that the steepest cusp always survives.
\end{abstract}

\section{Introduction}
The dynamical state of a stellar system is completely described by its `fine
grained' distribution function, $F(\bmath{x},\bmath{v},t)$, which refers to the
phase-space density at point $(\bmath{x},\bmath{v})$ and time $t$. The time
evolution of the distribution function is governed by a continuity equation,
known as the Vlasov or collisionless Boltzmann equation,
\begin{equation} \label{eq:cbe}
  {\mathrm{d}}_t F = \partial_t F + \bmath{v}\cdot \partial_{\bmath{x}} F
   - \partial_{\bmath{x}}\Phi\cdot \partial_{\bmath{v}} F = 0.
\end{equation}
Here, $\Phi(\bmath{x})$ denotes the gravitational potential, which for a
self-gravitating stellar system is given by the Poisson integral
\begin{equation} \label{eq:poisson}
  \Phi(\bmath{x},t) = -G \int
  \frac{F(\bmath{x}^\prime,\bmath{v},t)} {|\bmath{x}-\bmath{x}^\prime|} 
  {\mathrm{d}}^3\!\bmath{x}^\prime\,{\mathrm{d}}^3\!\bmath{v}.
\end{equation}
The main objective of galactic dynamics is to solve this system of equations.
This is a difficult task and most analytic work is restricted to stationary or
near-stationary solutions. For these a number of theoretical concepts have been
developed, such as Jeans' theorem and perturbation theory. Galactic dynamics far
from equilibrium on the other hand, such as in a galaxy merger or collapse, are
almost entirely treated with $N$-body simulations, i.e.\ numerical solutions of
equations (\ref{eq:cbe}) and (\ref{eq:poisson}).

As emphasised already by \cite{Henon1964} and \cite{LyndenBell1967}, the
constancy of $F(\bmath{x},\bmath{v},t)$ ensured by the collisionless Boltzmann
equation~(\ref{eq:cbe}) is of little practical use in non-equilibrium
situations, because of \emph{mixing}. Phase-space elements of high density are
stretched out and folded with elements of low density, very much as cream
stirred into coffee. As for this example, the elements become ever thinner until
any measurement of $F(\bmath{x},\bmath{v},t)$ becomes impossible. In other
words, the finite resolution of the system breaks the validity of the continuum
limit. In such a case, the system is better described by a local average of $F$,
known as the \emph{coarse-grained} distribution function
$\Bar{F}(\bmath{x},\bmath{v},t)$. In fact, any measurement can only (hope to)
recover this local average.

There are important differences between mixing in a collisionless system such as
a galaxy, and a collisional system such as a gas, where mixing is driven by
short-range interactions. In galaxies, strong forms of mixing are caused by
non-local large-scale dynamics and occur only away from equilibrium but
generally promote equilibrium. Hence, mixing is never complete in the sense of
convergence to a maximum-entropy state -- in fact, it can be shown that such a
state does not exist \citep*{TremaineHenonLyndenBell1986}. A mild version of
mixing is phase-mixing or similar `weak mixing' processes, caused by secular
evolution of stellar systems (for instance, the merging of regular orbits into a
sea of chaos mixes their phase-space densities \citep{MerrittValluri1996}).
Stronger forms of mixing are `chaotic mixing' or `violent relaxation', which is
driven by large-scale fluctuations of the gravitational potential.

\begin{figure*}
  \centerline{\hfil
    \resizebox{63mm}{!}{\includegraphics{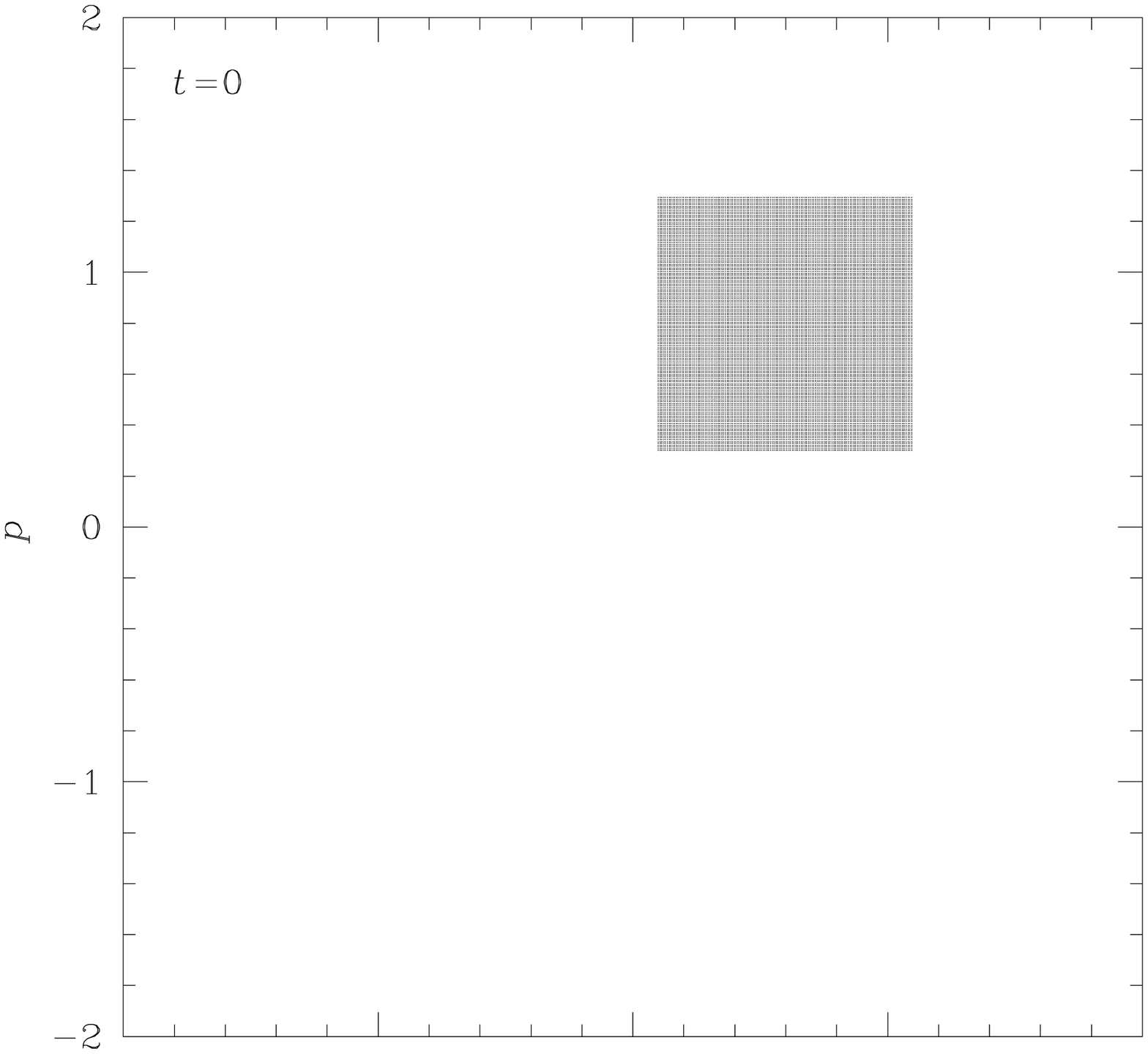}}\hspace*{-8mm}
    \resizebox{63mm}{!}{\includegraphics{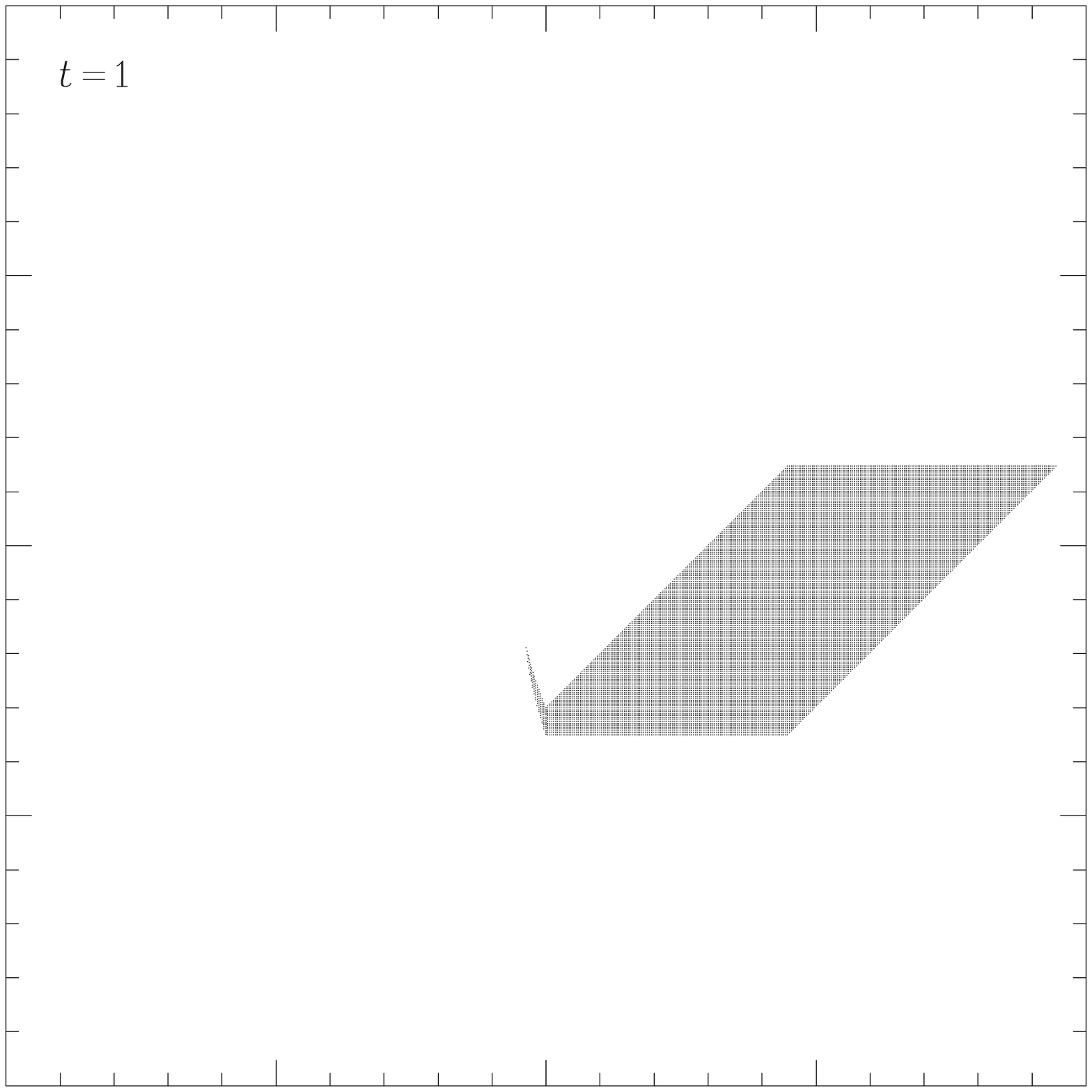}}\hspace*{-8mm}
    \resizebox{63mm}{!}{\includegraphics{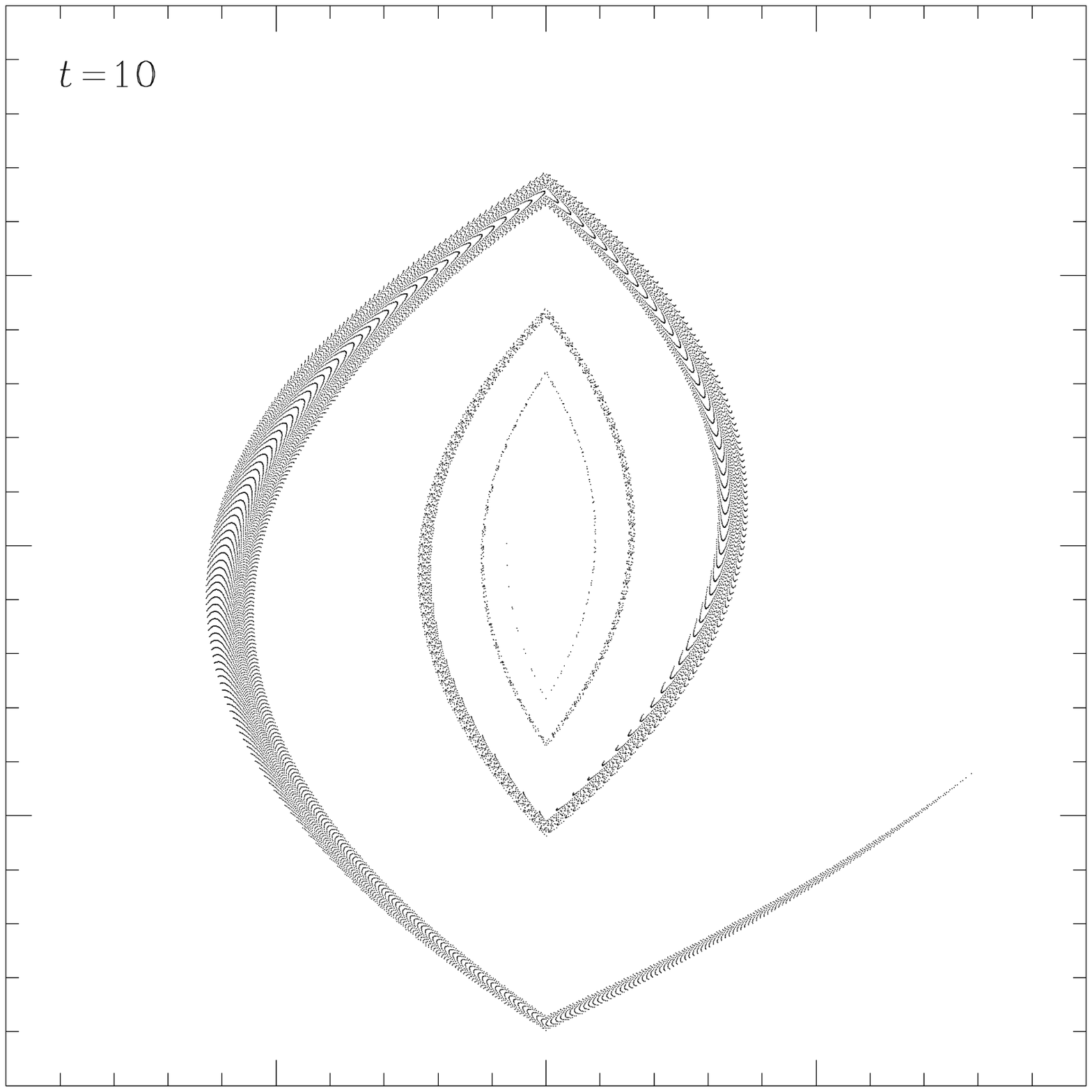}}\hfil
  }\vspace*{-8mm}
  \centerline{\hfil
    \resizebox{63mm}{!}{\includegraphics{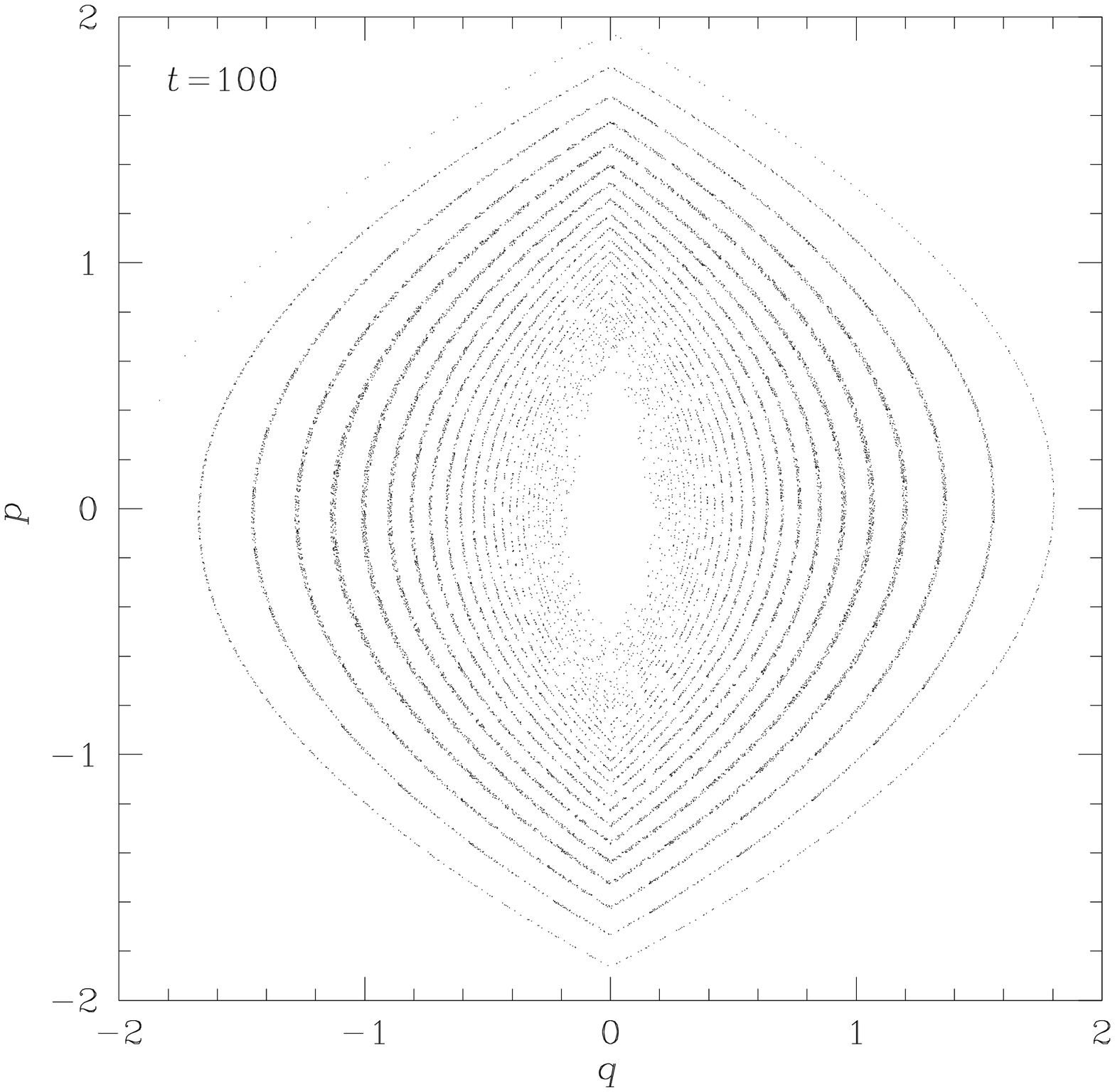}}\hspace*{-8mm}
    \resizebox{63mm}{!}{\includegraphics{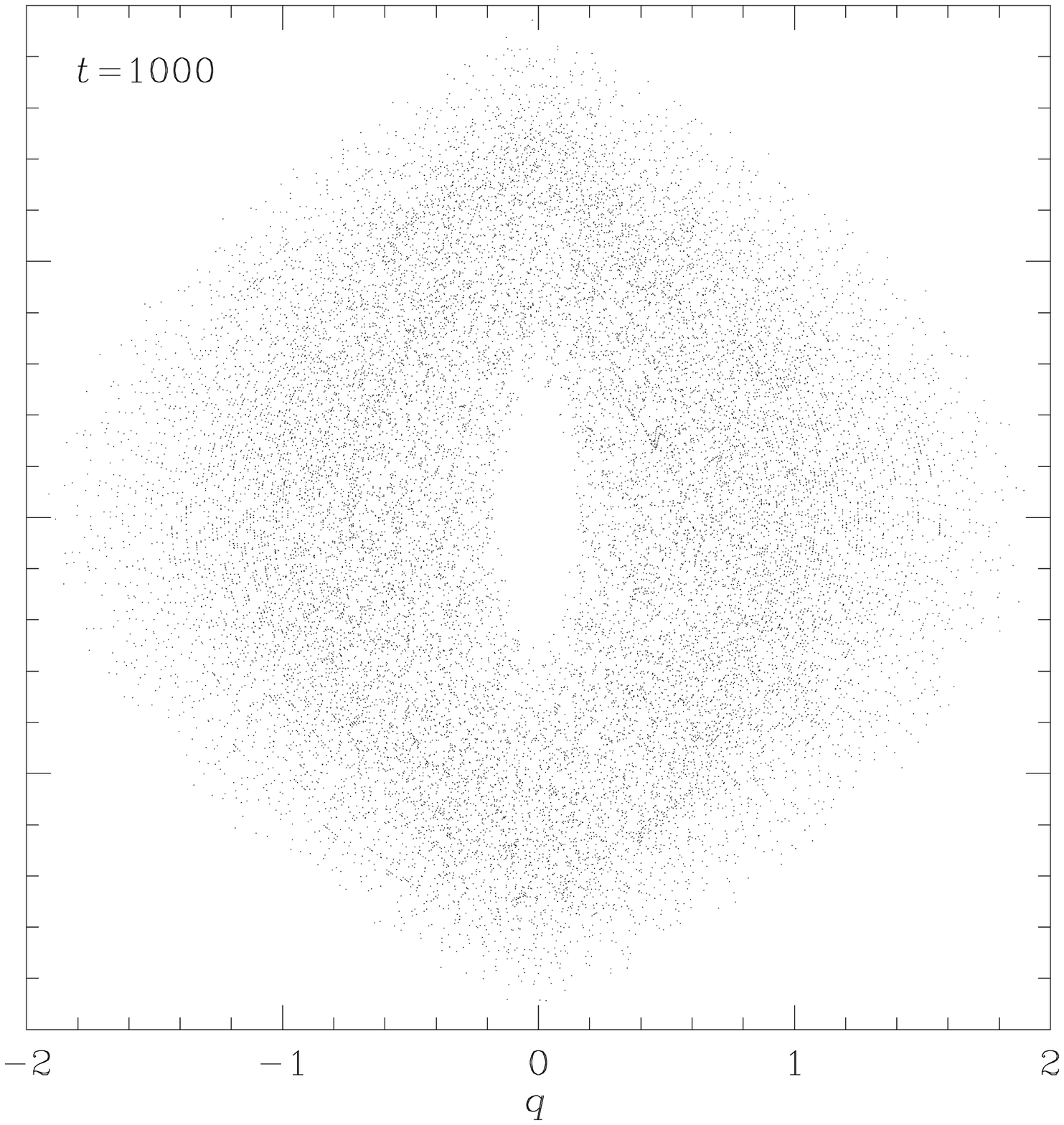}}\hspace*{-8mm}
    \resizebox{63mm}{!}{\includegraphics{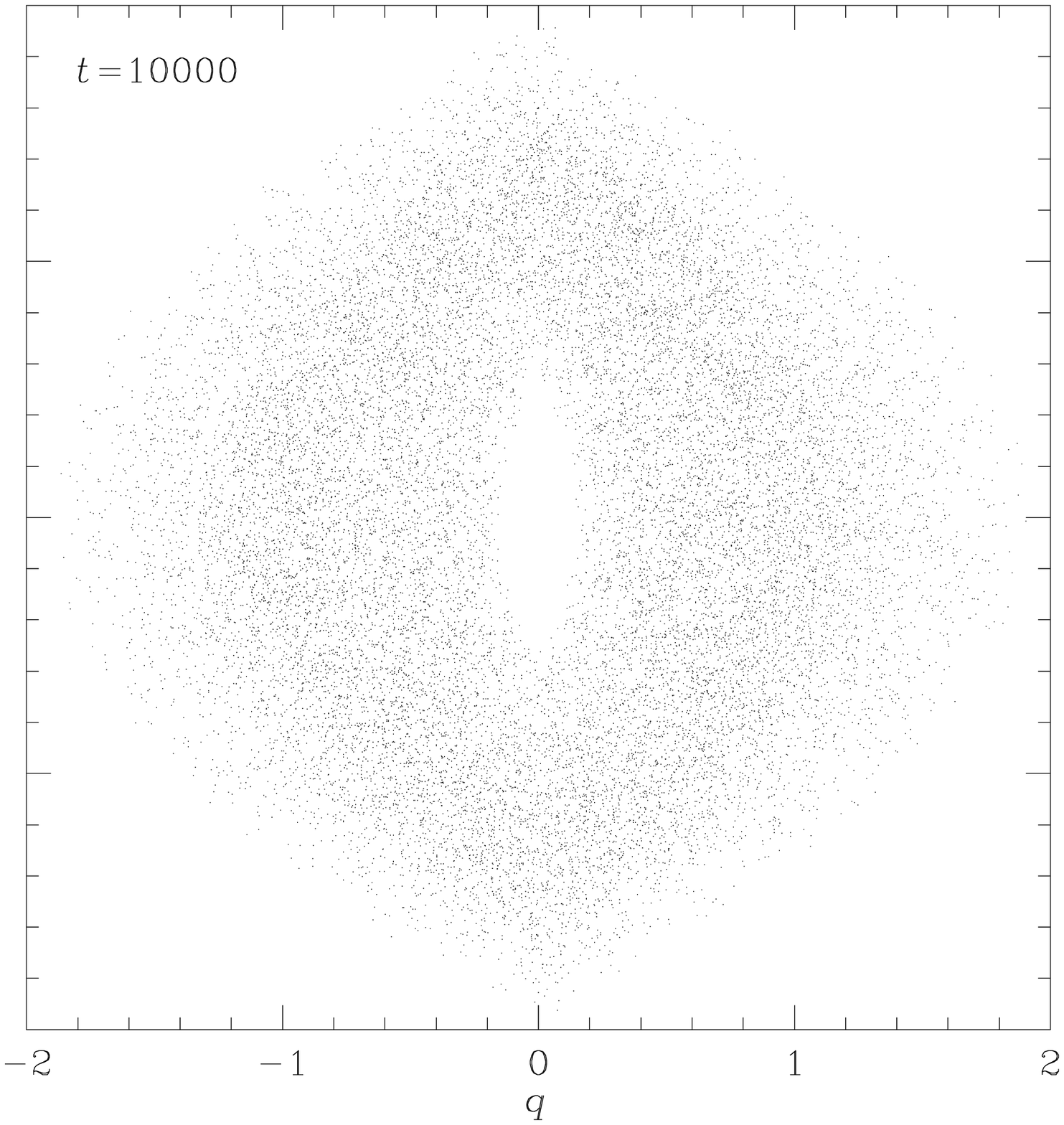}}\hfil
  }
  \caption{
    \label{fig:mixing}
    Demonstration of phase mixing. Shown is the evolution of a bunch of
    $\sim10^5$ phase-space points in the 1D Hamiltonian $H=p^2/2+|q|$ of a point
    mass in 1D gravity. The fine-grained distribution function is either equal
    to one (black) or zero (white), but at late times, a smooth distribution
    appears. }
\end{figure*}
A simple example of mixing is presented in Figure~\ref{fig:mixing}: volumes of
$F=1$ (black) get stretched out (e.g.\ at $t=100$) and folded together with
volumes of $F=0$ (white) until any distinction is barred by the finite
resolution of any observation. So, while at $t\ga1000$ in this example the
fine-grained distribution function $F(\bmath{x},\bmath{v},t)$ displays a very
complex pattern, equilibrium is reached in the coarse-grained sense, since
$\partial_t\Bar{F}=0$.

Unfortunately, $\Bar{F}$ does not obey a simple continuity equation, such as
(\ref{eq:cbe}), and hence describing its evolution is a considerable problem.
\cite{LyndenBell1967} derived a distribution function for the end-state of
violent relaxation assuming the conservation of phase-space volumes of given
density according to equation~(\ref{eq:cbe}). However, mixing does not conserve
volumes of fixed density in the coarse-grained sense \citep[e.g.][]{Mathur1988}
and, not surprisingly, the resulting theory is inconsistent
\citep{AradLyndenBell2005}. A similar attempt by \cite{Nakamura2000} suffers
from the same deficiency \citep{AradLyndenBell2005}. Another approach to the
dynamics of violent relaxation was taken by \cite{Chavanis1998} in deriving a
time evolution equation for $\Bar{F}$. While this is a promising attempt, its
practicality is limited and unlikely to surpass that of $N$-body simulations.

An obvious constraint on the evolution of $\Bar{F}(\bmath{x},\bmath{v},t)$ is
that its maximum value $\Bar{F}_{\mathrm{max}}$ cannot increase. While this is
applicable only if $\Bar{F}(\bmath{x},\bmath{v})$ is initially bounded, a much
stronger constraint on $\Bar{F}(\bmath{x},\bmath{v},t)$ is provided by a
\emph{mixing theorem}, a relation between the properties of
$\Bar{F}(\bmath{x},\bmath{v})$ before and after a mixing process.
\cite*{TremaineHenonLyndenBell1986} considered $H$-functionals of
$\Bar{F}(\bmath{x},\bmath{v},t)$, which are defined as\footnote{The traditional
  definition in statistical mechanics differs by a sign.}
\begin{equation} \label{eq:H}
  H[\Bar{F}] = - \int C\left(\Bar{F}(\bmath{x},\bmath{v})\right)\,
  {\mathrm{d}}^3\!\bmath{x}\,{\mathrm{d}}^3\!\bmath{v}
\end{equation}
where $C$ is a convex function with $C(0)=0$.
\citeauthor{TremaineHenonLyndenBell1986} \citep[see also][]{Tolman1938} showed
that coarse-graining always increases $H$-functionals and concluded that mixing
generally results in an increase of $H[\Bar{F}]$, a result known as the
`$H$-theorem'\footnote{\label{fn:coarse} This latter step, however, has been
  shown to be conceptually incorrect if one allows for arbitrary ways of
  coarse-graining \citep{Soker1996}.  Indeed, counter-examples to the mixing
  theorem can easily be constructed by tuning the coarse-graining
  \citep{Kandrup1987, Sridhar1987}.  It seems that the problem arises from
  problems with the very concept of coarse-graining, which is lacking precise
  pre-conditions or even a precise definition. This does, however, not imply
  that a statement like the $H$-theorem cannot generally be made or even that
  mixing was unimportant for stellar dynamics. I postpone a more detailed
  discussion of these issues to section~\ref{sec:coarse}.}.

According to \cite{TremaineHenonLyndenBell1986}, a function $\Bar{F}_2$ is
called \emph{more mixed} than $\Bar{F}_1$, if for all $H$-functionals
$H[\Bar{F}_2]\ge H[\Bar{F}_1]$. In particular, if $\Bar{F}_2$ originates from
$\Bar{F}_1$ by mixing, $\Bar{F}_2$ is more mixed than $\Bar{F}_1$. The
$H$-functional for $C(f)=f\ln f$ is equal to ($k_{\mathrm{B}}$ times) the
entropy and even increases for collisional systems.
\citeauthor{TremaineHenonLyndenBell1986} prove a mixing theorem stating that
\emph{$\Bar{F}_2$ is more mixed than $\Bar{F}_1$ if and only if
  $\mathcal{M}_2(V)\le \mathcal{M}_1(V)$ for all $V$}, where the function
$\mathcal{M}(V)$ is defined in terms of the cumulative volume and mass
\begin{eqnarray}
  \label{eq:V} V(f) &\equiv& \int_{\Bar{F}(\bmath{x},\bmath{v})>f}
  {\mathrm{d}}^3\!\bmath{x}\,{\mathrm{d}}^3\!\bmath{v}\\
  \label{eq:M} M(f) &\equiv& \int_{\Bar{F}(\bmath{x},\bmath{v})>f}
  \Bar{F}(\bmath{x},\bmath{v})\,
  {\mathrm{d}}^3\!\bmath{x}\,{\mathrm{d}}^3\!\bmath{v}\,
\end{eqnarray}
via $\mathcal{M}(V(f))\equiv M(f)$. This is a stronger statement than that of
the increase of entropy, since it holds for all values of $V$.  Unfortunately,
the usability of this theorem is restricted by the complicated definition of
$\mathcal{M}(V)$, which evades a simple interpretation and manipulation. For
instance, $\mathcal{M}(V)$ is not additive and, if $V(f)$ is not invertible, it
is not well defined, as is the case for the initial state of the example in
Fig.~\ref{fig:mixing}, where $V=1$ for $f\le1$ and $V=0$ for $f>1$ (known as the
`water-bag model').

The purpose of this paper is to present in section~\ref{sec:mix} a novel
approach to mixing which is conceptually different from that of
\citeauthor{TremaineHenonLyndenBell1986} and avoids their conceptual problems.
This leads to a novel form of the mixing theorem in terms of a new concept, the
excess-mass function, which is simple to apply and easy to interprete. I discuss
the relation to \citeauthor{TremaineHenonLyndenBell1986}'s work and the
controversy about it in section~\ref{sec:coarse}. In section~\ref{sec:ex},
simple examples are given and the asymptotic behaviour at small and large values
of $\Bar{F}$ corresponding, respectively, to large and small radii are
considered. These are applied in section~\ref{sec:merger} to the merging of
cusped galaxies. Finally, section~\ref{sec:Nbody} discusses the applicability to
$N$-body simulations and section~\ref{sec:conclude} concludes.

\section{Mixing} \label{sec:mix}

In order to simplify the following discussion, it is worth introducing the
`volume distribution function' \citep[e.g.][]{TremaineHenonLyndenBell1986}
\begin{equation} \label{eq:v}
  v(f) = \int\delta\left(\Bar{F}(\bmath{x},\bmath{v})-f\right)\,
  {\mathrm{d}}^3\!\bmath{x}\,{\mathrm{d}}^3\!\bmath{v},
\end{equation}
which refers to the phase-space volume at which
$\Bar{F}(\bmath{x},\bmath{v})=f$. Using $v(f)$, we can re-write the cumulative
mass and volume and any $H$-functional as
\begin{eqnarray}
  \label{eq:V:v} V(f)       &=&\phantom{-}
  \int_f^\infty v(\phi)\,{\mathrm{d}}\phi,\\
  \label{eq:M:v} M(f)       &=&\phantom{-}
  \int_f^\infty \phi\,v(\phi)\,{\mathrm{d}}\phi,\\
  \label{eq:H:v} H[\Bar{F}] &=& -
  \int_0^\infty v(\phi)\,C(\phi)\,{\mathrm{d}}\phi.
\end{eqnarray}
Note that $v(f)$, $V(f)$, $M(f)$, and $\mathcal{M}(V)$ are functionals of
$\Bar{F}(\bmath{x},\bmath{v})$ as well as functions of their parameter $f$ or
$V$. We can now model mixing directly as an operation on phase-space volumes.

\subsection{Infinitesimal mixing events} \label{sec:mix:elem}
The process of mixing and subsequent coarse-graining of the distribution
function can be described as sequence of infinitesimal mixing events in which a
phase-space element with infinitesimal volume ${\mathrm{d}} V_{\mathrm{l}}$ and
density $f_{\mathrm{l}}$ mixes completely with another volume ${\mathrm{d}}
V_{\mathrm{h}}$ having density $f_{\mathrm{h}}\ge f_{\mathrm{l}}$. Because of
conservation of mass and of phase-space volume, the resulting element has volume
${\mathrm{d}} V_{\mathrm{l}}+{\mathrm{d}} V_{\mathrm{h}}$ and density
\begin{equation} \label{eq:f:mix}
  f_{\mathrm{m}} = \frac{{\mathrm{d}} V_{\mathrm{l}}\,f_{\mathrm{l}} 
    +{\mathrm{d}} V_{\mathrm{h}}\,f_{\mathrm{h}}}
  {{\mathrm{d}} V_{\mathrm{l}} + {\mathrm{d}} V_{\mathrm{h}}}
\end{equation}
\citep{Mathur1988}. The change in the volume distribution function due to such
an event is
\begin{equation} \label{eq:v:mix}
  {\mathrm{d}} v(f) = 
  ({\mathrm{d}} V_{\mathrm{l}} + {\mathrm{d}} V_{\mathrm{h}})\,
  \delta(f-f_{\mathrm{m}})
  - {\mathrm{d}} V_{\mathrm{l}}\, \delta(f-f_{\mathrm{l}})
  - {\mathrm{d}} V_{\mathrm{h}}\, \delta(f-f_{\mathrm{h}})
\end{equation}
Mixing events with $f_{\mathrm{l}}=f_{\mathrm{h}}$ do not affect $v(f)$ and
hence may be called `adiabatic'.

\begin{figure}
  \centerline{\resizebox{\columnwidth}{!}{\includegraphics{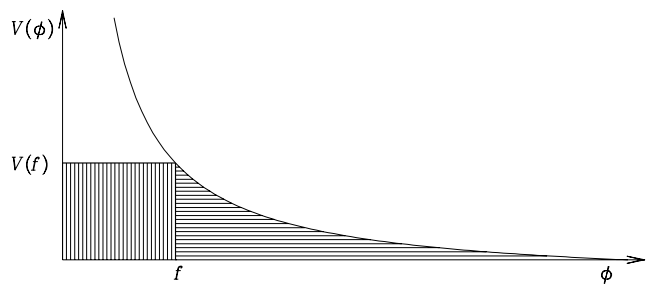}}}
  \caption{\label{fig:Df:def}
    In a plot of cumulative volume $V(\phi)$ vs.\ phase-space density $\phi$,
    the excess-mass $D(f)$ is given by the horizontally shaded region
    (equation~\ref{eq:Df:V}), which equals $M(f)$ (total shaded region) minus
    $fV(f)$ (vertically shaded; equation~\ref{eq:Df:MV}).}
\end{figure}

\subsection{A lemma on mixing} \label{sec:mix:lemma}
Consider the following function
\begin{equation} \label{eq:D:def}
  D(f) \equiv \int\limits_{\Bar{F}(\bmath{x},\bmath{v}) > f} 
  \left(\Bar{F}(\bmath{x},\bmath{v})-f\right)\,
  {\mathrm{d}}^3\!\bmath{x}\,{\mathrm{d}}^3\!\bmath{v},
\end{equation}
which may be re-written as
\begin{eqnarray}
  \label{eq:Df:v} D(f)
  &=& \int_f^\infty (\phi - f)\,v(\phi)\,{\mathrm{d}}\phi \\
  \label{eq:Df:V}
  &=& \int_f^\infty V(\phi)\,{\mathrm{d}}\phi,\\
  \label{eq:Df:MV}
  &=& M(f) - f\,V(f).
\end{eqnarray}
As is obvious from these relations, the \emph{excess-mass function} $D(f)$
refers to the excess mass due to values of $\Bar{F}>f$ (see also
Fig.~\ref{fig:Df:def}).

\paragraph*{Mixing lemma.} \emph{Mixing of phase-space volumes where
  $\Bar{F}(\bmath{x},\bmath{v})<f$ with volumes where
  $\Bar{F}(\bmath{x},\bmath{v})>f$ decreases $D(f)$; other mixing processes
  leave $D(f)$ unchanged.}

\bigskip
\noindent \emph{Proof}: First, consider an infinitesimal mixing event.
The changes it imposes on $D(f)$ are easily found from equations
(\ref{eq:v:mix}) and (\ref{eq:Df:v}):
\begin{equation} \label{eq:deltaDf}
  {\mathrm{d}} D(f) = \left\{
    \begin{array}{l@{\hspace{4mm}}l@{\hspace{4mm}}lllll}
      0             & 
      \mbox{for} &     &   &  f & < & f_{\mathrm{l}},\\
      -(f-f_{\mathrm{l}})\,{\mathrm{d}} V_{\mathrm{l}} & 
      \mbox{for} & f_{\mathrm{l}} & < &  f & < & f_{\mathrm{m}},\\
      -(f_{\mathrm{h}}-f)\,{\mathrm{d}} V_{\mathrm{h}} & 
      \mbox{for} & f_{\mathrm{m}} &\le&  f & < & f_{\mathrm{h}},\\
      0             & 
      \mbox{for} & f_{\mathrm{h}} & < &  f.     
    \end{array}
  \right.
\end{equation}
Thus, ${\mathrm{d}} D(f)<0$ for $f_{\mathrm{l}}<f<f_{\mathrm{h}}$ and 0
otherwise. Since the whole mixing process is a sequence of infinitesimal mixing
events, the change in $D(f)$ is the integral over many infinitesimal changes
${\mathrm{d}} D(f)$ and the lemma follows.

The largest change of $D(f)$ due to an infinitesimal mixing event occurs at
$f{=}f_{\mathrm{m}}$ and is ${\mathrm{d}} D(f_{\mathrm{m}}) = -
|f_{\mathrm{h}}-f_{\mathrm{l}}|{\mathrm{d}} V_{\mathrm{l}}\,{\mathrm{d}}
V_{\mathrm{h}}/({\mathrm{d}} V_{\mathrm{l}}+{\mathrm{d}} V_{\mathrm{h}})$.

\subsection{Further properties of the excess-mass function} 
\label{sec:mix:Dprop}
Apart from the relations (\ref{eq:Df:v}), (\ref{eq:Df:MV}), and
(\ref{eq:Df:V}), 
the function $D(f)$ has the following properties. First,
\begin{eqnarray}
  \label{eq:dDf}
  D^\prime(f) &=& -V(f),\\
  \label{eq:ddDf}
  D^{\prime\prime}(f) &=& \phantom{-} v(f).
\end{eqnarray}
Since both $v(f)$ and $V(f)$ are non-negative, this implies that $D(f)$ is
non-negative and monotonically declining with everywhere non-negative curvature.

Second, since in order for $M(f\to\infty)$ not to diverge $f^2v(f)\to0$ as
$f\to\infty$, and because of equation (\ref{eq:ddDf}),
\begin{equation}
  \lim_{f\to\infty} D(f) = 0.
\end{equation}

Third, for a system with finite mass,
\begin{equation}
  D(0) = M_{\mathrm{total}}.
\end{equation}

Fourth, changes in $D(f)$ are related to a change of the entropy $S$ via
\begin{equation} \label{eq:entropy}
  \Delta S =  - k_{\mathrm{B}} \int_0^\infty
  \frac{\Delta D(f)}f\,{\mathrm{d}} f,
\end{equation}
which becomes obvious at the end of the next section.

Finally, the combined excess-mass function of several \emph{disjoint} systems
(whose distribution functions do not overlap) is simply given by the sum of the
individual excess-mass functions. In general, i.e.\ for partially overlapping
systems, the excess-mass function is \emph{super-additive}:
\begin{equation}
  D_{1+2}(f) \ge D_1(f)+D_2(f).
\end{equation}
This is directly related to the sub-additivity of the entropy
\citep[e.g.][]{Wehrl1978} and follows from the definition (\ref{eq:D:def}) of
$D(f)$ and the fact that $\Bar{F}_{1+2}=\Bar{F}_1+\Bar{F}_2$.

\subsection{A mixing theorem} \label{sec:mix:theorem}
The above lemma is closely related to a statement made by \cite{Mathur1988}. In
fact, his function $P(f)$, for which he only gives the second derivative, is
identical to the change in $D(f)$ induced by mixing and his equation (7) is
equivalent to my (\ref{eq:deltaDf}).

The relation to the theorem given by \cite{TremaineHenonLyndenBell1986} and
outlined in the beginning of this section is more subtle. In the proof of their
theorem, \citeauthor{TremaineHenonLyndenBell1986} construct the function $D(f)$,
because it actually is (the negative of) an $H$-functional of $\Bar{F}$. In
fact, as pointed out by \cite{Mathur1988}, $D(f)$ and $\mathcal{M}(V)$ are
related by a Legendre transformation\footnote{\citeauthor{Mathur1988} failed to
  derive $D(f)$ itself, but based his statement on equation (\ref{eq:ddDf}),
  which he used as definition; also he strangely considered negative values for
  $f$.}, as is obvious from equation (\ref{eq:Df:MV}). In particular,
$\mathcal{M}^\prime(V)=f$, $\mathcal{M}^{\prime\prime}(V)=1/v(f)$, and
$D^\prime$ (equation~\ref{eq:dDf}) is the (negative of the) inverse of
$\mathcal{M}^\prime$, by definition of a Legendre transform. This is directly
related to the fact that
\begin{equation} \label{eq:dD:dM}
  \left(\frac{\partial D}{\partial\tau}\right)_f = 
  \left(\frac{\partial\mathcal{M}}{\partial\tau}\right)_V
\end{equation}
where $\tau$ is a time-like variable describing the evolution due to mixing.
Equation (\ref{eq:dD:dM}) in conjunction with the theorem given by
\cite{TremaineHenonLyndenBell1986} implies the following alternative form of the
mixing theorem.

\paragraph*{Mixing theorem.} \emph{The distribution function
  $\Bar{F}_2(\bmath{x},\bmath{v})$ is more mixed than
  $\Bar{F}_1(\bmath{x},\bmath{v})$ if and only if $D_2(f)\le D_1(f)$ for all
  $f$.}

\bigskip
\noindent \emph{Proof}: Suppose $\Bar{F}_2$ is more mixed than $\Bar{F}_1$.
Then $D_2(f)\le D_1(f)$, since $D(f)$ is the negative of a $H$-functional of
$\Bar{F}(\bmath{x},\bmath{v})$ with the convex function
\begin{equation}
  C(\Bar{F}) = \left\{
    \begin{array}{l@{\hspace{4mm}}l}
      0         & \mbox{for $\quad\Bar{F}\le f$},\\
      \Bar{F}-f & \mbox{for $\quad\Bar{F} >  f$}.
    \end{array}
    \right.
\end{equation}

Conversely, suppose $D_2(f)\le D_1(f)$ for all $f$. From equation
(\ref{eq:H:v}),
\begin{equation}
  H_2-H_1 = - \int_0^\infty (v_2-v_1) C\,{\mathrm{d}} f.
\end{equation}
Integrating by parts and using $V^\prime=-v$ yields
\begin{equation}
  H_2-H_1 = (V_2-V_1) C\Big|_0^\infty 
  - \int_0^\infty (V_2-V_1) C^\prime\,{\mathrm{d}} f.
\end{equation}
The first term on the right-hand side vanishes, and integrating the second term
by parts using (\ref{eq:dDf}) gives
\begin{equation} \label{eq:theorem}
  H_2-H_1 = (D_2-D_1) C^\prime\Big|_0^\infty
  - \int_0^\infty (D_2-D_1) C^{\prime\prime}\,{\mathrm{d}} f.
\end{equation}
Again the first term on the right-hand side vanishes; the second term is
non-negative, since $D_2\le D_1$ by assumption and $C^{\prime\prime}\ge0$ by
definition of convexity. Hence $H_2 \ge H_1$, which completes the proof.

Together with the above lemma, this theorem is another proof of the $H$-theorem
(mixing increases or preserves but never decreases any $H$-functional).

Since the entropy $S$ is the $H$-functional with $C(f)=f\ln f$, the
relation~(\ref{eq:entropy}) between changes in entropy and $D(f)$ follows
directly from equation~(\ref{eq:theorem}).

\subsection{Diluting phase-space density} \label{sec:mix:dilute}
The theorem above is more useful than the equivalent theorem by
\cite{TremaineHenonLyndenBell1986} because the function $D(f)$ is easier to
comprehend and manipulate than $\mathcal{M}(V)$. In particular the additivity of
$D(f)$ is of great value. However, the lemma of \S\ref{sec:mix:lemma} is of even
larger practical significance, because it allows us to relate the change in
$D(f)$ directly to mixing events `across' $\Bar{F}=f$.

The total change of $D(f)$ in a mixing process may be obtained by integrating
the infinitesimal change (\ref{eq:deltaDf}) over a function which specifies for
each pair $(f_{\mathrm{l}},f_{\mathrm{h}})$ how much phase space at
$f_{\mathrm{l}}$ mixes with how much phase space at $f_{\mathrm{h}}$
\citep{Mathur1988}. However, it is not clear how such a function may be
obtained; moreover, in the end the information contained in this function is
reduced to the one-dimensional change in $D(f)$. One may instead assume a simple
form for this function, resulting in simple mixing models

For instance, one may assume that all of $v(f)$ gets mixed with an empty volume
of size $\alpha(f)v(f)$. This complete `mixing with air' simply \emph{dilutes}
the phase-space density $\Bar{F}\to\Bar{F}/[1+\alpha(\Bar{F})]$ and gives
\begin{equation} \label{eq:Df:dilute}
  D_{\mathrm{final}}(f) = D_{\mathrm{initial}}([1+\alpha(f)] f).
\end{equation}
The \emph{dilution function} $\alpha(f)$ is always well-defined and can be
measured directly from $N$-body experiments, by estimating $D(f)$ before and
after a violent mixing process. Essentially $\alpha(f)$ gives the equivalent
amount of mixing with air necessary to generate a certain evolution of $D(f)$.

\section{Coarse-Graining} \label{sec:coarse}
\subsection{Conceptual problems} \label{sec:coarse:problem}
As mentioned in footnote~\ref{fn:coarse} above, the $H$-theorem of
\cite{TremaineHenonLyndenBell1986} has met immediate rejection
\citep{Dejonghe1987, Kandrup1987, Sridhar1987}. These authors provided simple
counter-examples of non-mixing systems whose $H$-functionals are not conserved
or even decreasing and pointed to the following conceptual problem in the
argumentation.  \citeauthor{TremaineHenonLyndenBell1986} have actually only
proven that $H$-functionals increase as consequence of coarse-graining:
$H[\Bar{F}(\bmath{x},\bmath{v})]\ge H[F(\bmath{x},\bmath{v})]$ independent of
the actual dynamics or indeed mixing. From that, they argued that initially
$\Bar{F}=F$ (which can be guaranteed by definition of the arrow of time,
Tremaine, private communication), but $\Bar{F}\neq F$ at a later time $t$, and
hence $H[\Bar{F}(0)]\le H[\Bar{F}(t)]$. However, this argument does not
guarantee that $H[\Bar{F}(0)]\le H[\Bar{F}(t_1)]\le H[\Bar{F}(t)]$ at
intermediate times $t_1$.

Unfortunately, this controversy undermined the whole subject of mixing in
stellar dynamics. Some sceptics argue that the fine-grained distribution
function never suffers information loss and hence that the $H$-theorem is a pure
artifact of coarse-graining (implying that the entropy of a collisionless
stellar system is constant). This argument, however, relies on the infinite
resolution of the fine-grained distribution function and ignores the fact that
no stellar system can support infinite resolution. The fine-grained distribution
function and the CBE only give an \emph{approximative} description of
collisionless stellar dynamics \citep[e.g.][]{Dejonghe1987}. In the presence of
mixing, the continuum limit, on which this approximation rests, becomes invalid.
Mixing is an irreversible process as information about the state prior to mixing
is lost, representing a true entropy increase in the sense understood by
Boltzmann \citep{Merritt1999}.

The conceptual problems related to the $H$-theorem originate from the fact that
the details of and requirements for the coarse-graining operation are not
specified and hence are usually considered unimportant. Many authors consider a
static coarse-graining operation, such as averaging over time-independent macro
cells or convolution with a window function. For such ways of coarse-graining
\cite{Soker1996} showed that $H[\Bar{F}]$ does not obey a
$H$-theorem\footnote{As a simple example consider $F$ to be the sum of two
  $\delta$-functions. If the two points are close enough to be within one macro
  cell, $\Bar{F}_{\mathrm{max}}$ is twice as large than otherwise.}. The reason
is easily understood when considering a non-mixing non-equilibrium (e.g.\ 
periodic) system. Since the system evolves, the effectively resolved mass per
fixed macro cell evolves too, so that $H[\Bar{F}]$ is not necessarily conserved.

\subsection{A novel interpretation of coarse-graining} \label{sec:coarse:cond}
In this situation, it is instructive to consider the proof of the $H$-theorem
from the previous section. Unlike \citeauthor{TremaineHenonLyndenBell1986}'s
proof, it does not employ coarse-graining with finite macro cells. Rather mixing
is described directly as (integral over) averaging of infinitesimal phase-space
volumes. In this description, the astrophysical process of mixing (and the
resulting loss of information) is accounted for, in our description of the
system, by a local averaging, the coarse-graining. Thus, mixing and
coarse-graining are intimately related and the latter must not be considered
arbitrary.

This is directly related to the interpretation of the coarse-grained
distribution function $\Bar{F}$. Traditionally, $\Bar{F}$ is introduced, because
its fine-grained pendant $F$ does not tend to equilibrium, but undergoes ever
stronger small-scale fluctuations \citep[e.g.][]{Chavanis1998}. In this picture,
$\Bar{F}$ gives an otherwise unspecified, finite-resolution representation of
the system.  As already mentioned, the fine-grained fluctuations of $F$ will
eventually break the validity of the continuum approximation, i.e.\ below some
level, these fluctuations are artificial and not representative of the actual
stellar system.

These arguments suggest the interpretation of $\Bar{F}$ as our \emph{best
  possible description} of the stellar system, avoiding the artifacts of its
fine-grained counterpart. In this interpretation, coarse-graining must meet the
following conditions.
\begin{enumerate}
\item[1.] \label{cond:one}
  In the limit of a system with infinite resolution: $\Bar{F}\to F$.
\item[2.] \label{cond:two}
  $\Bar{F}$ must be a faithful representation of the system.
\item[3.] \label{cond:three}
  In the absence of mixing: ${\mathrm{d}}_t\Bar{F}=0$.
\end{enumerate}
The first two conditions imply that coarse-graining must be local. The second
condition ensures that coarse-graining only deletes information in $F$ but not
in our description of the stellar system, in particular $\Bar{\Bar{F}}=\Bar{F}$.
This means that coarse-graining is done on the local resolution scale of the
system itself, which seems the most natural scale to use, but excludes static
coarse-graining. In particular, any moment of the stellar system must agree
(within its statistical uncertainty) with the corresponding moment of $\Bar{F}$.
Finally, the last condition guarantees that $\Bar{F}$ is altered by mixing only,
i.e.\ under ordinary non-mixing circumstances all information about the system
is preserved in $\Bar{F}$. This immediately warrants the $H$-theorem.

It is not clear, whether and how these conditions can be met in a practical
implementation. However, even if we could only generate an approximation to this
ideal, the above conditions and the underpinning interpretation were still
valuable.  For instance, we would allow for an approximation error
$\Hat{\Bar{F}}-\Bar{F}$ (with $\Hat{\cdot}$ denoting approximation), which in
turn might result in spurious but accountable violations of the $H$-theorem.
Clearly, a more detailed investigation of these issues is beyond the scope of
this paper.

We should stress that the idea of $\Bar{F}(\bmath{x},\bmath{v},t)$ being the
best possible description of the stellar system is not necessarily consistent
with other approaches. For instance, \citeauthor{Chavanis1998}
(\citeyear{Chavanis1998}, see also \citealt{ChavanisBouchet2005}) consider
$\Bar{F}$ to contain a truly reduced information and, hence, the fluctuations of
$F$ to be (at least partially) real rather than entirely artificial. It may be
possible to reconcile this with our ideas by altering the above conditions to
allow for a arbitrary resolution (in terms of the mass or number of stars per
resolution element) without spoiling the $H$-theorem -- however, then
${\mathrm{d}}_t\Bar{F}\neq0$ because of the forces generated by the
fluctuations \citep{Chavanis1998}.

\section{Examples} \label{sec:ex}
\subsection{Asymptotics at large \boldmath$f$} \label{sec:ex:high}
\begin{figure}
  \centerline{\resizebox{\columnwidth}{!}{\includegraphics{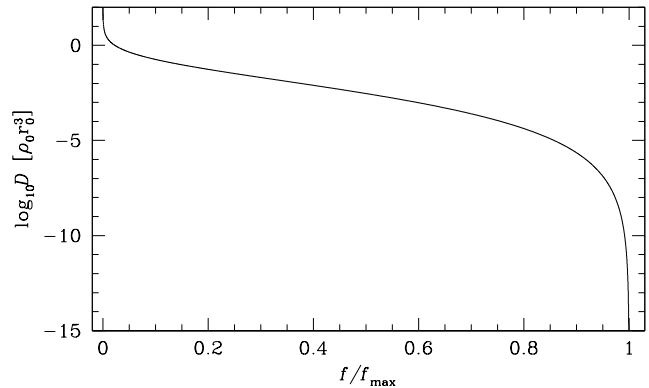}}}
  \caption{
    \label{fig:isothermal}
    The excess-mass function $D(f)$ for the (non-singular) isothermal sphere.}
\end{figure}

\subsubsection{Density cores: limited phase-space densities}
\label{sec:ex:high:core}
Let us consider the (non-singular) isothermal sphere
\citep[e.g.,][]{BinneyTremaine1987}.  I have numerically obtained the potential
$\Phi(r)$ as well as the phase-space volume $g(E)$ at constant energy. The
volume distribution function is then given by $g(E) / |{\mathrm{d}} F /
{\mathrm{d}} E|$ and $D(f)$ can be computed using equation~(\ref{eq:Df:v}).  The
result is plotted in Figure~\ref{fig:isothermal}. The mass of the isothermal
sphere is infinite and hence $D\to\infty$ as $f\to0$. The maximum 
phase-space density is $f_{\mathrm{max}} = 2\rho_0/[9(2\pi\sigma)^{3/2}]$, where
$\sigma$ and $\rho_0$ denote the velocity dispersion and central density,
respectively. At $f\approx f_{\mathrm{max}}$, the excess-mass function decays
like $D\propto (f_{\mathrm{max}}-f)^4$.

\stepcounter{footnote}
\subsubsection{Density cusps: unlimited phase-space
  densities$^\arabic{footnote}$}
\label{sec:ex:high:cusp} \addtocounter{footnote}{-1}
Let us consider a self-gravitating stellar system whose density at small radii
is given by a power law in radius, corresponding to a \emph{density
  cusp}\footnote{For any real system, the resolution is of course finite and
  hence, the density limited. However, we assume here that the resolution is
  high enough for the asymptotic limit to be useful over a range of densities.}
\begin{equation} \label{eq:cusp:rho}
  \rho(\bmath{x}) = \rho_0\,\varrho(\Hat{\bmath{x}})\, x^{-\gamma}
\end{equation}
with $\Hat{\bmath{x}}\equiv\bmath{x}/|\bmath{x}|$ the unit vector in direction
of $\bmath{x}$ and the dimensionless radius $x\equiv|\bmath{x}|/r_{\mathrm{u}}$,
where $r_{\mathrm{u}}$ is the unit of length. The parameters $\rho_0$ and
$0<\gamma<3$ are, respectively, a density normalisation and cusp strength. The
dimensionless and continuous function $\varrho(\Hat{\bmath{x}})$ determines the
shape of surfaces of equal density. I proceed by assuming that the distribution
function is of the form
\begin{equation} \label{eq:cusp:df}
  \Bar{F} = \rho_0 v_0^{-3} h(\bmath{X}) \tilde f(E)
\end{equation}
with the constant $v_0$ given in equation~(\ref{eq:vzero}). Here
$E=\bmath{v}^2/2+\Phi(\bmath{x})$ is the energy, while $\bmath{X} = \bmath{X}
(\bmath{x}, \bmath{v})$ denotes a set of \emph{scale-invariant} integrals of
motion. Scale invariance in this case means that
\begin{equation} \label{eq:scalefree}
  \bmath{X}(\bmath{x},\bmath{v}) = \bmath{X}(a\bmath{x},a^{1-\gamma/2}\bmath{v})
\end{equation}
for any dimensionless scale factor $a$. Examples for scale-invariant
properties of stellar orbits are the eccentricity and ratios between orbital
frequencies or actions. After some algebra (see appendix~\ref{sec:cusp}), the
excess-mass of the self-gravitating cusp function is found to be
\begin{equation} \label{eq:cusp:Df}
  D(f)=\mathrm{D}_{\gamma h}\,r_{\mathrm{u}}^3\rho_0\,
  (f^2\,G^3\rho_0r_{\mathrm{u}}^6)^{-\frac{3-\gamma}{6-\gamma}}.
\end{equation}
Here, $\mathrm{D}_{\gamma h}$ is a dimensionless constant given in
equation~(\ref{eq:cusp:Dgh}), which for the spherical ($\varrho\equiv1$) and
isotropic ($h\equiv1$) case reduces to a simple expression (see
appendix~\ref{sec:cusp}).

The exponent in $D\propto f^{-2(3-\gamma)/(6-\gamma)}$ varies only between $1$
for $\gamma\to0$ and 0 for $\gamma\to3$ and decreases with increasing $\gamma$,
i.e.\ a steeper cusp is less mixed than a shallower one.  Moreover, this
asymptotic behaviour of $D(f)$ depends only on the cusp strength and not on the
details of the density contours or the distribution function, as long as it is
scale-invariant.

For a non-self-gravitating system with density $\propto r^{-\gamma}$ immersed in
a gravitational field generated by an overall mass density $\rho\propto
r^{-\beta}$ with $\beta\ge\gamma$, one finds by a similar analysis
\begin{equation} \label{eq:non_self}
  D\propto f^{-2(3-\gamma)/(6+2\gamma-3\beta)}.
\end{equation}

\subsubsection{Stellar systems dominated by a super-massive black hole}
The case of a stellar system whose dynamics is dominated by a super-massive
black hole corresponds to $\beta=3$ in equation (\ref{eq:non_self}), i.e.\ gives
$D\propto f^{-2(3-\gamma)/(2\gamma-3)}$. Note that the phase-space density of
such a system is unlimited only for $\gamma>3/2$. The exponents in this relation
vary between $\infty$ for $\gamma\to3/2$ and $0$ for $\gamma\to3$.  Thus, again
steeper cusps are less mixed, but the differences are much more pronounced than
for self-gravitating cusps.

\subsection{Asymptotics at small \boldmath$f$} \label{sec:ex:low}
Next, consider a stellar system of finite mass. For simplicity, I consider the
case of spherical density only. The potential in the outer parts is dominated by
the monopole, i.e.\ $\Phi=-GM/r$, while the density is assumed to be of the form
$\rho=\rho_0(r/r_{\mathrm{u}})^{-\eta}$ with $\eta>3$ for the mass not to
diverge at $r\to\infty$. A distribution function of the form $\Bar{F}\propto
L^{-2\beta} \tilde{f}(E)$ always generates a constant Binney anisotropy
$\beta\equiv1-\sigma_\theta^2/\sigma_r^2$ \citep{Cuddeford1991}. For the case
considered here, this gives
\begin{equation} \label{eq:low:fEX}
  \Bar{F} = \rho_0\,(GM/r_{\mathrm{u}})^{-3/2}\,
  \mathrm{f}_{\eta\beta}\, X^{-\beta}\,
  \mathcal{E}^{\eta-3/2},
\end{equation}
where $\mathcal{E}=-r_{\mathrm{u}} E/GM$, $X\equiv L^2/L^2_{\mathrm{circ}}(E) =
1-\epsilon^2$ (orbital circularity), and $\mathrm{f}_{\eta\beta}^{-1}=2^{3/2}\pi
B(\frac{1}{2},1-\beta)B(\frac{3}{2}-\beta,\eta-\frac{1}{2}-\beta)$. The
phase-space volume at fixed $(E,X)$
\begin{equation} \label{eq:low:gEX}
  g(E,X) = \sqrt{2GMr_{\mathrm{u}}^5}\,\pi^3\mathcal{E}^{-5/2}
\end{equation}
is independent of $X$. From these, one can obtain $v(f)$ and
\begin{equation} \label{eq:low:D}
  D(f) = M_{\mathrm{D}} - \rho_0r_{\mathrm{u}}^3\,\mathrm{D}_{\eta\beta}\,
  \left(\frac{f^2G^3M^3}{r_{\mathrm{u}}^3\rho_0^2}
  \right)^{\frac{\eta-3}{2\eta-3}}
\end{equation}
with $M_{\mathrm{D}}\le M$ the total mass of the density component considered
and
\begin{equation}
  \mathrm{D}_{\eta\beta} = 
  \frac{2^{1/2}\,\pi^3(2\eta-3)^2}{3(\eta-3)(2\eta-3-3\beta)}\,
  \mathrm{f}_{\eta\beta}^{3/(2\eta-3)}.
\end{equation}
Hence, again the asymptotic is independent of details like the orbital
anisotropy.

\begin{figure}
  \centerline{\resizebox{\columnwidth}{!}{\includegraphics{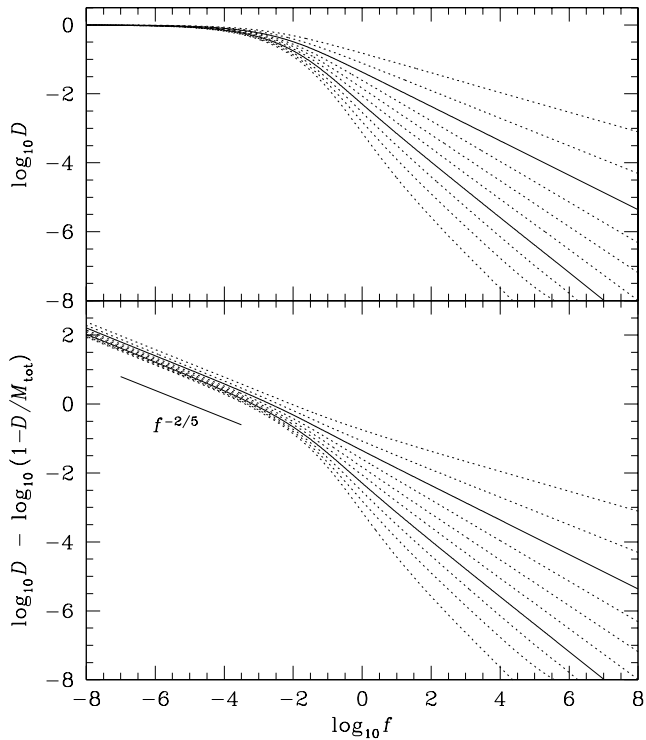}}}
  \caption{
    \label{fig:gamma}
    \textbf{Top}: The excess-mass function $D(f)$ for spherical and isotropic
    $\gamma$ models, which have density $\rho\propto
    r^{-\gamma}(r+a)^{\gamma-4}$. The curves correspond to values of $\gamma$
    from 0.25 (\emph{lowest}) to 2.5 (\emph{uppermost}) spaced by 0.25 with
    $\gamma=1$ and $\gamma=2$ plotted \emph{solid}. Units correspond to $G$, $a$
    and total mass equal to unity. \textbf{Bottom}: The same, but plotting
    $D/(1-D/M_{\mathrm{tot}})$, which distinguishes the models at
    $f\to0$ and demonstrates that in this limit they indeed behave as
    equation~(\ref{eq:low:D}) predicts for $\eta=4$.}
\end{figure}

\subsection{Excess-mass functions of the \boldmath$\gamma$ models}
\label{sec:ex:models}
Figure~\ref{fig:gamma} shows the excess-mass function of the spherical
$\gamma$-models \citep[][with $M$ and $a$ denoting total mass and scale
radius]{Dehnen1993,TremaineEtal1994}
\begin{equation}
  \rho = \frac{3-\gamma}{2\pi}\frac{Ma}{r^\gamma(r+a)^{4-\gamma}}
\end{equation}
with isotropic velocity distribution. Evidently, $D_1 < D_2$ for $\gamma_1 <
\gamma_2$, thus $\gamma$-models are less mixed with increasing $\gamma$.  The
bottom panel shows $D/(1-D/M_{\mathrm{tot}})$ and enables to better distinguish
between the excess-mass functions at $f\to0$. The line shows the asymptotic
slope predicted by equation~(\ref{eq:low:D}).

\section{Application: merging Cusps} \label{sec:merger}
As application, consider the merging of several cusped galaxies or dark-matter
haloes. Because of its additivity the combined $D(f)$ prior to the merger is
equal to the sum of those of the progenitors. When equilibrium has been
re-established after the merger, $D_{\mathrm{r}}(f)$ of the merger remnant must
satisfy
\begin{equation} \label{eq:D:merged}
  D_{\mathrm{r}}(f) \le D_{\Sigma}(f) \equiv \sum_i D_i(f).
\end{equation}

\subsection{Constraints on the cusp strength} \label{sec:merger:gamma}
Let us first consider the combined excess-mass function of the progenitors on
the right-hand side of (\ref{eq:D:merged}). Each $D_i(f)$ is of the
form~(\ref{eq:cusp:Df}): $D_i\propto f^{-2(3-\gamma_i)/(6-\gamma_i)}$.  Thus, at
sufficiently large phase-space densities $D_{\Sigma}$ will be dominated by the
steepest progenitor cusp. Next suppose the remnant also forms a scale-free cusp,
such that its excess-mass function $D_{\mathrm{r}}\propto
f^{-2(3-\gamma_{\mathrm{r}})/(6-\gamma_{\mathrm{r}})}$.  Then for condition
(\ref{eq:D:merged}) to be satisfied $\gamma_{\mathrm{r}}\le\max_i\{\gamma_i\}$.
Thus, \emph{the remnant cusp cannot be steeper than any of the progenitor
  cusps}.

In other words, steeper cusps are less mixed than shallower cusps (in the limit
$f\to\infty$) regardless of details such as the shape of the density contours or
the distribution of orbital shapes (orbital anisotropy), as long as these are
the same for all radii and/or energies (scale freedom). This implies that by
virtue of the mixing theorem mergers cannot produce cusps steeper than those
already present in their progenitors. Conversely, a remnant cusp shallower than
the steepest of its progenitors would require an arbitrarily large dilution
$\alpha$ of the distribution function as $f\to\infty$. This, while not
impossible, seems highly implausible, which strongly suggests that the remnant
cusp should not be shallower than the steepest of its progenitors.  Together
with the above mixing constraint, this means that the maximum cusp strength is
conserved when merging collisionless stellar systems.

\subsection{Constraints on the cusp mass} \label{sec:merger:mass}
Let me exemplify the merging of two equal and cusped galaxies a little more.  I
assume that the remnant has the same scale-invariant structure and cusp strength
$\gamma$ as its progenitors, but different density normalisation,
$\rho_{0\mathrm{r}}\neq\rho_{0\mathrm{p}}$.  Under these circumstances, the
dilution function $\alpha(f)$ of equation~(\ref{eq:Df:dilute}) is constant (in
the asymptotic limit considered). Together with equation~(\ref{eq:cusp:Df}) and
the additivity, this yields
\begin{equation} \label{eq:merge:alpha}
  \rho_{0\mathrm{r}} = \rho_{0\mathrm{p}}\,
  2^{-\frac{6-\gamma}{3}}\,(1+\alpha)^{-\frac{3}{2(3-\gamma)}}.
\end{equation}
Thus, for $\rho_{0\mathrm{r}}=2\rho_{0\mathrm{p}}$ (i.e.\ the remnant cusp
containing the sum of the progenitor-cusp masses), the dilution fraction has to
be $\alpha=\sqrt{2}-1\approx41\%$ independent of $\gamma$. For the remnant cusp
to be equally massive as either of its progenitors,
$\alpha=2^{(6-\gamma)/2(3-\gamma)}-1$, which evaluates to 1 for $\gamma=0$, to
$\approx138\%$ for $\gamma=1$, to $\approx183\%$ for $\gamma=1.5$, and $3$ for
$\gamma=2$. Thus, if $\alpha$ is not strongly dependent on $\gamma$,
relation~(\ref{eq:merge:alpha}) suggests that remnants of steep-cusp mergers
have more massive cusps, compared to their progenitors, than remnants of
shallow-cusp mergers. However, since steeper cusps generate stronger tides, one
would indeed expect $\alpha$ to depend on $\gamma$ in a sense opposing the above
trend.

\section{Application to \boldmath$N$-body simulations}
\label{sec:Nbody}
One of the motivations of this study was the hope to use the excess-mass
function as a diagnostic tool in the interpretation and validation of $N$-body
experiments. To this end it is necessary to estimate phase-space densities from
$N$-body data. \cite*{AradDekelKlypin2004} and \cite{AscasibarBinney2005} have
demonstrated in two pioneering studies that this can in principle be done, even
though it is a difficult task, because of (i) the vastness of 6D phase space and
(ii) the lack of a metric.

Once estimates $\Hat{\Bar{F}}_i$ for $\Bar{F}$ at the phase-space positions of
the bodies have been obtained, the excess-mass function may be estimated as
\begin{equation} \label{eq:D:estimate}
  \Hat{D}(f) = \sum_{\Hat{\Bar{F}}_i>f} m_i - f
              \sum_{\Hat{\Bar{F}}_i>f} \frac{m_i}{\Hat{\Bar{F}}_i},
\end{equation}
where $m_i$ denotes the mass of the $i$th body. A serious problem in this game
is that the coarse-graining scale used in the estimation of $\Bar{F}$ may well
be larger than the scales still resolved by the system. As discussed in
section~\ref{sec:coarse}, this approximation error may result in unwanted
surprises (such as $\Hat{D}$ increasing with time).

\begin{figure}
  \centerline{\resizebox{\columnwidth}{!}{\includegraphics{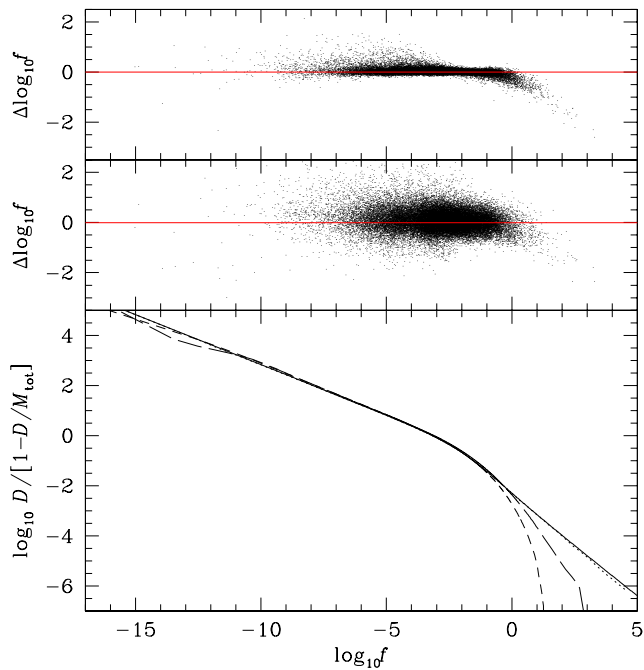}}}
  \caption{
    \label{fig:FiEstAS}
    Assessing the fitness of \textsc{FiEstAS} for estimating $D(f)$.
    \textbf{Top}: error in the \textsc{FiEstAS} estimates $\Hat{\Bar{F}}_i$ for
    $\Bar{F}$ from $N=10^6$ points drawn from a Hernquist
    (\citeyear{Hernquist1990}) sphere ($\gamma=1$ in Fig.~\ref{fig:gamma})
    plotted versus their true phase-space density (only every 25th body
    plotted).  \textbf{Middle}: the same, but using \textsc{FiEstAS} without
    smoothing.  \textbf{Bottom}: Excess-mass function of the same model
    (\emph{solid}), and estimates for $D(f)$ using
    equation~(\ref{eq:D:estimate}) and the true fine-grained $F$ of $N=10^6$
    points (\emph{dotted}), or the \textsc{FiEstAS} estimates $\Hat{\Bar{F}}_i$
    for the same points with (\emph{long dashed}) and without smoothing
    (\emph{short dashed}).}
\end{figure}

Figure~\ref{fig:FiEstAS} demonstrates these problems when applying the
phase-space density estimator \textsc{FiEstAS} of \cite{AscasibarBinney2005} to
$10^6$ points drawn from a Hernquist (\citeyear{Hernquist1990}) sphere. The top
two panels show that the estimated density can be up to two orders of magnitude
wrong and is systematically too large for $f\la0.01$, while high densities are
truncated, in particular when smoothing is used (top panel). The bottom panel
shows the estimate $\Hat{D}(f)$ obtained from equation~(\ref{eq:D:estimate}) and
the \textsc{FiEstAS} estimated densities with (\emph{short dashed}) and without
(\emph{long dashed}) smoothing. A comparison with the true $D(f)$ (\emph{solid})
shows that the estimated $D(f)$ is only slightly too large at intermediate $f$,
but is seriously in error at large $f$ (and also at very small $f$), in
particular when smoothing was used. These errors are obviously related to the
underestimation of large phase-space densities. This underestimation occurs
at values for $\Bar{F}$ below those expected from the finite resolution with
$N=10^6$, since $M(f)/M_{\mathrm{tot}}=10^{-6}$ at $f\approx10^{4.7}$ and
$10^{-5}$ at $f\approx10^{3.5}$.

Also plotted are the estimates from equation~(\ref{eq:D:estimate}) using the
true fine-grained phase-space densities of $10^6$ bodies (\emph{dotted}). These
can hardly be distinguished from the true $D(f)$, indicating that
equation~(\ref{eq:D:estimate}) gives a fairly good estimator provided the
estimates $\Hat{\Bar{F}}_i$ are good.

These results suggest that, in order to resolve the asymptotic behaviour of
$D(f)$ at $f\to\infty$, at least $\sim10^7$ points are required with this
technique. Clearly, there must be ways to improve the situation. Apart from
improvements in the existing technique, one may exploit that
${\mathrm{d}}_t\Bar{F}=0$ (assuming no mixing is going on) in order to constrain
the possible values for $\Hat{\Bar{F}}_i$.

\section{Summary and Conclusion} \label{sec:conclude}
A stellar system out of equilibrium is driven towards equilibrium by way of
\emph{mixing} its phase-space densities in a process of violent relaxation. As a
consequence the concept of the fine-grained distribution function
$F(\bmath{x},\bmath{v})$ is ill-suited to understand non-equilibrium stellar
dynamics. Instead, the system is better described in terms of its coarse-grained
distribution function $\Bar{F}(\bmath{x},\bmath{v})$. Mixing of phase-space
elements changes $\Bar{F}$ in such a way that the \emph{excess-mass function}
\[
D(f) \equiv \int\limits_{\Bar{F}(\bmath{x},\bmath{v}) > {f}}
\left(\Bar{F}(\bmath{x},\bmath{v})-{f}\right)\,
{\mathrm{d}}^3\!\bmath{x}\,{\mathrm{d}}^3\!\bmath{v}
\]
decreases (\emph{mixing lemma}, section~\ref{sec:mix:lemma}), equivalent to a
statement by \cite{Mathur1988}. In fact, only events which mix densities $>f$
with densities $<f$ decrease $D(f)$. This lemma may be considered an extension
of the well-known maximum phase-space density argument to all density values.
$D(f)$ measures the excess mass due to phase-space densities higher than $f$ and
its decrease is directly related to entropy increase, see
equation~(\ref{eq:entropy}). A useful property of $D(f)$ is its additivity: the
excess-mass function of the combination of disjoint stellar systems (which do
not overlap in phase space) is simply the sum of the individual $D(f)$.

In section~\ref{sec:mix:theorem}, I prove a novel form of the \emph{mixing
  theorem} \citep*{TremaineHenonLyndenBell1986}, stating that $D(f)$ decreases
if and only if any $H$-functional of the distribution function increases. My
lemma together with this theorem is an alternative proof of the $H$-theorem (the
increase of $H$-functionals due to mixing), avoiding some conceptual problems
associated with allowing arbitrary coarse-graining.

In section~\ref{sec:coarse}, the importance of details of the coarse-graining
operator for the validity of the $H$-theorem are discussed. It is argued that
the conceptual problems of \citeauthor{TremaineHenonLyndenBell1986}'s proof of
this theorem can be avoided by requiring appropriate coarse-graining. In
particular, I propose an interpretation of the coarse-grained distribution
function $\Bar{F}$ as the best possible description of the stellar system. In
this interpretation, the astrophysical process of mixing is directly described
by coarse-graining and in the absence of mixing ${\mathrm{d}}_t\Bar{F}=0$, such
that the $H$-theorem is guaranteed.

In section~\ref{sec:ex}, $D(f)$ for some simple spherical equilibria is given
and its asymptotic behaviour at small and large $f$ considered.  For equilibria
with a self-gravitating scale-invariant density cusp ($\rho\propto
r^{-\gamma}$), the asymptotic behaviour at $f\to\infty$ is $D\propto
f^{-2(3-\gamma)/(6-\gamma)}$ \emph{independent} of the shape of the density
contours and details of the distribution function. This remarkable property
together with the additivity and the mixing theorem allowed me in
section~\ref{sec:merger} to prove that a merger remnant cannot have a density
cusp steeper than any of its progenitors.  Assuming that mixing during the
merger does not become ever stronger at higher values of $\Bar{F}$ (which cannot
be strictly excluded, but appears highly implausible), one can show that the
maximum cusp strength is conserved, i.e.\ the remnant cusp has strength $\gamma$
equal to the maximum of its progenitors.

Clearly, the decreasing nature of the excess-mass function is not restricted to
galactic dynamics, but applicable to any collisionless system undergoing mixing.
For instance, the inequality constraint used by \citeauthor{YuTremaine2002}
(\citeyear{YuTremaine2002}, eq.~33) to describe the evolution of the population
of super-massive black holes is essentially equivalent to the mixing lemma. In
this case, merging of super-massive black holes mixes the distribution of their
properties.

\section*{acknowledgement}
Theoretical Astrophysics at Leicester is supported by a PPARC rolling grant. The
author thanks several members of this group, as well as Scott Tremaine, for
helpful discussions and Yago Ascasibar for providing his code \textsc{FiEstAS}.


\begin{thebibliography}{}
  
\bibitem[\protect\citeauthoryear{{Arad}, {Dekel} \& {Klypin}}{{Arad}
    et~al.}{2004}]{AradDekelKlypin2004} {Arad} I., {Dekel} A., {Klypin} A.,
  2004, MNRAS, 353, 15
  
\bibitem[\protect\citeauthoryear{{Arad} \& {Lynden-Bell}}{{Arad} \&
    {Lynden-Bell}}{2005}]{AradLyndenBell2005} {Arad} I., {Lynden-Bell} D., 2005,
  MNRAS, submitted (astro-ph/0409728)
  
\bibitem[\protect\citeauthoryear{{Ascasibar} \& {Binney}}{{Ascasibar} \&
    {Binney}}{2005}]{AscasibarBinney2005} {Ascasibar} Y., {Binney} J., 2005,
  MNRAS, 356, 872
  
\bibitem[\protect\citeauthoryear{{Binney} \& {Tremaine}}{{Binney} \&
    {Tremaine}}{1987}]{BinneyTremaine1987} {Binney} J.~J., {Tremaine} S., 1987,
  {Galactic dynamics}.  Princeton, NJ, Princeton University Press
  
\bibitem[\protect\citeauthoryear{{Chavanis}}{{Chavanis}}{1998}]{Chavanis1998}
  {Chavanis} P.-H., 1998, MNRAS, 300, 981
  
\bibitem[\protect\citeauthoryear{{Chavanis} \& {Bouchet}}{{Chavanis} \&
    {Bouchet}}{2005}]{ChavanisBouchet2005} {Chavanis} P.~H., {Bouchet} F., 2005,
  A\&A, 430, 771
  
\bibitem[\protect\citeauthoryear{{Cuddeford}}{{Cuddeford}}{1991}]{Cuddeford1991}
  {Cuddeford} P., 1991, MNRAS, 253, 414

\bibitem[\protect\citeauthoryear{{Dehnen}}{{Dehnen}}{1993}]{Dehnen1993}
  {Dehnen} W., 1993, MNRAS, 265, 250
  
\bibitem[\protect\citeauthoryear{{Dejonghe}}{{Dejonghe}}{1987}]{Dejonghe1987}
  {Dejonghe} H., 1987, ApJ, 320, 477

\bibitem[\protect\citeauthoryear{{H{\' e}non}}{{H{\' e}non}}{1964}]{Henon1964}
  {H{\' e}non} M., 1964, Annales d'Astrophysique, 27, 83

\bibitem[\protect\citeauthoryear{{Hernquist}}{{Hernquist}}{1990}]{Hernquist1990}
  {Hernquist} L., 1990, ApJ, 356, 359

\bibitem[\protect\citeauthoryear{{Kandrup}}{{Kandrup}}{1987}]{Kandrup1987}
  {Kandrup} H.~E., 1987, MNRAS, 225, 995

\bibitem[\protect\citeauthoryear{{Lynden-Bell}}{{Lynden-Bell}}{1967}]
  {LyndenBell1967} {Lynden-Bell} D., 1967, MNRAS, 136, 101
  
\bibitem[\protect\citeauthoryear{{Mathur}}{{Mathur}}{1988}]{Mathur1988} {Mathur}
  S.~D., 1988, MNRAS, 231, 367
  
\bibitem[\protect\citeauthoryear{{Merritt}}{{Merritt}}{1999}]{Merritt1999}
  {Merritt} D., 1999, PASP, 111, 129
  
\bibitem[\protect\citeauthoryear{{Merritt} \& {Valluri}}{{Merritt} \&
    {Valluri}}{1996}]{MerrittValluri1996} {Merritt} D., {Valluri} M., 1996, ApJ,
  471, 82
  
\bibitem[\protect\citeauthoryear{{Nakamura}}{{Nakamura}}{2000}]{Nakamura2000}
  {Nakamura} T.~K., 2000, ApJ, 531, 739
  
\bibitem[\protect\citeauthoryear{{Soker}}{{Soker}}{1996}]{Soker1996} {Soker} N.,
  1996, ApJ, 457, 287
  
\bibitem[\protect\citeauthoryear{{Sridhar}}{{Sridhar}}{1987}]{Sridhar1987}
  {Sridhar} S., 1987, Journal of Astrophysics and Astronomy, 8, 257
  
\bibitem[\protect\citeauthoryear{{Tolman}}{{Tolman}}{1938}]{Tolman1938} {Tolman}
  R.~C., 1938, {The Principles of Statistical Mechanics}.  Clarendon Press,
  Oxford
  
\bibitem[\protect\citeauthoryear{{Tremaine}, {Henon} \&
    {Lynden-Bell}}{{Tremaine} et~al.}{1986}]{TremaineHenonLyndenBell1986}
  {Tremaine} S., {Henon} M., {Lynden-Bell} D., 1986, MNRAS, 219, 285
  
\bibitem[\protect\citeauthoryear{{Tremaine}, {Richstone}, {Byun}, {Dressler},
    {Faber}, {Grillmair}, {Kormendy} \& {Lauer}}{{Tremaine}
    et~al.}{1994}]{TremaineEtal1994} {Tremaine} S., {Richstone} D.~O., {Byun}
  Y., {Dressler} A., {Faber} S.~M., {Grillmair} C., {Kormendy} J., {Lauer}
  T.~R., 1994, AJ, 107, 634
  
\bibitem[\protect\citeauthoryear{{Wehrl}}{{Wehrl}}{1978}]{Wehrl1978} {Wehrl} A.,
  1978, Rev.\ Mod.\ Phys., 50, 221
  
\bibitem[\protect\citeauthoryear{{Yu} \& {Tremaine}}{{Yu} \&
    {Tremaine}}{2002}]{YuTremaine2002} {Yu} Q., {Tremaine} S., 2002, MNRAS, 335,
  965

\end{thebibliography}

\appendix
\section{\boldmath $D(\mbox{\lowercase{$f$}})$ for density cusps}
\label{sec:cusp}
Here, we derive the excess-mass function for stellar systems with a power-law
density (\ref{eq:cusp:rho}) and a scale-invariant distribution function of the
form~(\ref{eq:cusp:df}).

The gravitational potential generated by the density (\ref{eq:cusp:rho}) is
\begin{equation} \label{eq:cusp:Phi}
  \Phi(\bmath{x}) = v_0^2\, \times
  \left\{
    \begin{array}{l@{\hspace{4mm}}l}
      \displaystyle \frac{\psi(\Hat{\bmath{x}})}{2-\gamma}\,
      x^{2-\gamma}
      & \mbox{for $\quad \gamma\neq2$}, \\[2ex]
      \displaystyle \ln \psi(\Hat{\bmath{x}}) + \ln x
      & \mbox{for $\quad \gamma = 2$}
    \end{array}\right.
\end{equation}
with
\begin{equation} \label{eq:vzero}
  v_0^2\equiv \frac{4\,\pi\,G\rho_0r_{\mathrm{u}}^2}{3-\gamma},
\end{equation}
where $G$ denotes Newton's constant of gravity. Here, $\psi(\Hat{\bmath{x}})$ is
a dimensionless shape function which is uniquely determined by
$\varrho(\Hat{\bmath{x}})$ and $\gamma$ through Poisson's equation, giving
\begin{equation} \label{eq:Poisson:angle}
  \varrho = \left\{
    \begin{array}{l@{\hspace{4mm}}l}
      \displaystyle \Big(1+\frac{\Delta_\Omega}{(2-\gamma)(3-\gamma)}\Big)\,\psi
      & \mbox{for $\quad \gamma\neq2$},\\[2ex]
      \displaystyle (1+\Delta_\Omega)\ln\psi
      & \mbox{for $\quad \gamma = 2$}
    \end{array}\right.
\end{equation}
with $\Delta_\Omega =(\sin\theta)^{-1}\partial_\theta
\sin\theta\,\partial_\theta +(\sin\theta)^{-2}\partial^2_\varphi $ the angular
part of the Laplace operator. Naturally, for the sperical case
$\varrho\equiv1\equiv\psi$.

The assumed functional form~(\ref{eq:cusp:df}) for $\Bar{F}$ means that the
distribution of scale invariant orbital properties is the same for all energies
and is determined by the function $h(\bmath{X})$. For self-consistency,
$\Bar{F}$ must generate the density (\ref{eq:cusp:rho}), which uniquely
determines the energy dependence, yielding
\begin{equation} \label{eq:gen:f}
  \Bar{F} =\rho_0 v_0^{-3} \mathrm{f}_{\gamma h}\,h(\bmath{X})\,
  x_E^{-(6-\gamma)/2}
\end{equation}
with $\mathrm{f}_{\gamma h}$ a normalisation constant and
\begin{equation} \label{eq:xE}
  x_E = \left\{
    \begin{array}{l@{\hspace{4mm}}l}
      \displaystyle
      \Big((2-\gamma)\frac{E}{v_0^2}\Big)^{1/(2-\gamma)}
      & \mbox{for $\quad\gamma\neq2$},\\[1ex]
      \displaystyle
      \exp \frac{E}{v_0^2}
      & \mbox{for $\quad\gamma=2$}.
    \end{array}
  \right.
\end{equation}
Inserting this into the self-consistency constraint $\rho(\bmath{x}) =
\int\Bar{F}\, {\mathrm{d}}^3\!\bmath{v}$, one finds after some algebra
\begin{equation} \label{eq:gen:relation}
  \varrho(\Hat{\bmath{x}}) = \mathrm{f}_{\gamma h}
  \int{\mathrm{d}}^3\!\bmath{w} \, 
  h\big(\bmath{X}(r_{\mathrm{u}}\Hat{\bmath{x}}, v_0\bmath{w})\big)\,
  \zeta(\Hat{\bmath{x}},\bmath{w})^{-\frac{6-\gamma}{2}}
\end{equation}
with
\begin{equation} \label{eq:zeta}
  \zeta(\Hat{\bmath{x}},\bmath{w}) = \left\{
    \begin{array}{l@{\hspace{4mm}}l}
      \displaystyle
      \left(\frac{2-\gamma}{2}\bmath{w}^2 +
      \psi(\Hat{\bmath{x}})\right)^{\frac{1}{2-\gamma}}
      & \mbox{for $\quad\gamma\neq2$},\\[2ex]
      \displaystyle
      \mathrm{e}^{\bmath{w}^2/2}\,\psi(\Hat{\bmath{x}})
      & \mbox{for $\quad\gamma=2$}.
    \end{array}\right.
\end{equation}
The integral in equation~(\ref{eq:gen:relation}) is over all $\bmath{w}$-space
for $\gamma\le2$ and restricted to $\bmath{w}^2/2 \le
\psi(\Hat{\bmath{x}})/(\gamma-2)$ for $\gamma>2$.
Equation~(\ref{eq:gen:relation}) is the self-consistency constraint for the
function $h(\bmath{X})$ and determines the constant $\mathrm{f}_{\gamma h}$. For
the spherical case ($\varrho\equiv1$) with isotropic velocity distribution
($h\equiv1$) equation~(\ref{eq:gen:relation}) yields (with $B(x,y)$ the beta
function)
\begin{equation} \label{eq:iso:fg}
  \mathrm{f}_\gamma = \left\{
    \begin{array}{l@{\hspace{4mm}}l}
      \displaystyle
      \frac{(2-\gamma)^{3/2}}
      {2^{5/2}\,\pi\,B\big(\frac{3}{2},\frac{\gamma}{2-\gamma}\big)}
      & \mbox{for $\quad\gamma<2$},\\[3ex]
      \displaystyle
      \pi^{-3/2}
      & \mbox{for $\quad\gamma=2$},\\[1ex]
      \displaystyle
      \frac{(\gamma-2)^{3/2}}
      {2^{5/2}\,\pi\,B\big(\frac{3}{2},\frac{\gamma+2}{2(\gamma-2)}\big)}
      & \mbox{for $\quad\gamma>2$},
    \end{array}\right.
\end{equation}
which is continuous in $\gamma$.

Next, the phase-space volume $g(E,\bmath{X})$ at fixed $E$ and $\bmath{X}$ is
obtained by integrating $\delta(E-\bmath{v}^2/2-\Phi(\bmath{x}))
\delta(\bmath{X}_0{-}\bmath{X}(\bmath{x},\bmath{v}))$ over all phase space. By
exploiting the scale-invariance of $\bmath{X}$ one obtains after a little
algebra
\begin{equation}
  g(E,\bmath{X}) = r_{\mathrm{u}}^3 v_0\,
  \mathrm{g}_\gamma(\bmath{X})\,x_E^{(8-\gamma)/2}
\end{equation}
with
\begin{equation} \label{eq:gbX}
  \mathrm{g}_\gamma(\bmath{X}_0) = \int
  \frac{{\mathrm{d}}^2\!\Hat{\bmath{x}}\,{\mathrm{d}}^3\!\bmath{w}}
  {\zeta(\Hat{\bmath{x}},\bmath{w})^{\frac{3(4-\gamma)}{2}}}
  \,\delta\big(\bmath{X}_0 -
  \bmath{X}(r_{\mathrm{u}} \Hat{\bmath{x}}, v_0\bmath{w})\big),
\end{equation}
where $\int{\mathrm{d}}^2\!\Hat{\bmath{x}}$ denotes the integral over the
sphere, while the integral ${\mathrm{d}}^3\bmath{w}$ is over the same volume as
in equation~(\ref{eq:gen:relation}).  For the spherical case, the phase-space
volume at fixed energy $g(E)$ may be obtained by integrating $g(E,\bmath{X})$
over all $\bmath{X}$, which gives $g(E)=r_{\mathrm{u}}^3
v_0\,\mathrm{g}_\gamma\,x_E^{(8-\gamma)/2}$ with
\begin{equation} \label{eq:iso:gg}
  \mathrm{g}_\gamma = (4\pi)^2\,\sqrt{2}\,\left\{
    \begin{array}{l@{\hspace{4mm}}l}
      \displaystyle
      \frac{B\big(\frac{3}{2},\frac{3}{2-\gamma}\big)} {(2-\gamma)^{3/2}}
      & \mbox{for $\quad\gamma<2$},\\[3ex]
      \displaystyle
      \sqrt{3\pi}/18
      & \mbox{for $\quad\gamma=2$},\\[1ex]
      \displaystyle
      \frac{B\big(\frac{3}{2},\frac{8-\gamma}{2(\gamma-2)}\big)}
      {(\gamma-2)^{3/2}}
      & \mbox{for $\quad\gamma>2$},
    \end{array}\right.
\end{equation}
which is continuous in $\gamma$.

Now, one can compute the volume distribution function as
\begin{equation}
  v(f) = \int g(E,\bmath{X})\,\delta(f-\Bar{F}(E,\bmath{X}))\,
  {\mathrm{d}} E\,{\mathrm{d}}\bmath{X} 
\end{equation}
and derive the excess-mass function via equation (\ref{eq:Df:v}) to be of the
form~(\ref{eq:cusp:Df}) with the dimensionless constant
\begin{equation} \label{eq:cusp:Dgh}
  \mathrm{D}_{\gamma h} = \frac{(6-\gamma)\,
  \mathrm{f}_{\gamma h}^{\frac{3-\gamma}{6-\gamma}}}{3(3-\gamma)(4-\gamma)}
  \left(\frac{4\pi}{3-\gamma}\right)^{\frac{9-\gamma}{6-\gamma}}
  \int\mathrm{g}_\gamma(\bmath{X})\,
  h(\bmath{X})^{\frac{3(4-\gamma)}{6-\gamma}}{\mathrm{d}}\bmath{X}.
\end{equation}
For the spherical ($\varrho\equiv1$) and isotropic ($h\equiv1$) case, the
integral in this equation is just the constant $\mathrm{g}_\gamma$ of
equation~(\ref{eq:iso:gg}) and $\mathrm{f}_{\gamma h}$ is given by equation
(\ref{eq:iso:fg}).

\end{document}